Parker Solar Probe evidence for scattering of electrons in the young solar wind by narrowband whistler-mode waves

C. Cattell[1], A. Breneman[1], J. Dombeck[1], B. Short[1], J. Wygant[1], J. Halekas[2], Tony Case[6], J. C. Kasper[4,5], D. Larson[3], Mike Stevens[6], P. Whittesley[3], S. Bale[3,11], T. Dudok de Wit[7], K. Goodrich[3], R. MacDowall[8], M. Moncuquet[10], D. Malaspina[9], M. Pulupa[3]

1. School of Physics and Astronomy, University of Minnesota, 116 Church St. SE Minneapolis -mail:cattell@umn.edu

2. Department of Physics and Astronomy, University of Iowa, Iowa City, IA 52242, USA

3. Space Sciences Laboratory, University of California, Berkeley, CA 94720-7450, USA

4. BWX Technologies, Inc., Washington DC 20002, USA

5. Climate and Space Sciences and Engineering, University of Michigan, Ann Arbor, MI 48109, USA

6. Smithsonian Astrophysical Observatory, Cambridge, MA 02138 USA

7. LPC2E, CNRS, CNES and University of Orléans, Orléans, France

8. Solar System Exploration Division, NASA/Goddard Space Flight Center, Greenbelt, MD, 20771

9. Laboratory for Atmospheric and Space Physics, University of Colorado, Boulder, CO 80303, USA

10. LESIA, Observatoire de Paris, Université PSL, CNRS, Sorbonne Université, Université de Paris, 5 place Jules Janssen, 92195 Meudon,France

11. Department of Physics, University of California, Berkeley, Berkeley, CA 94708

Abstract: Observations of plasma waves by the Fields Suite and of electrons by the Solar Wind Electrons Alphas and Protons Investigation on Parker Solar Probe provide strong evidence for pitch angle scattering of strahl-energy electrons by narrowband whistler-mode waves at radial distances less than ~0.3 AU. We present two example intervals of a few hours each that include eight waveform captures with whistler-mode waves and 26 representative electron distributions that are examined in detail. Two were narrow; seventeen were clearly broadened, and eight were very broad. The two with narrow strahl occurred when there were either no whistlers or very intermittent low amplitude waves. Six of the eight broadest distributions were associated with intense, long duration waves. Approximately half of the observed electron distributions have features consistent with an energy dependent scattering mechanism, as would be expected from interactions with narrowband waves. A comparison of the wave power in the whistler-mode frequency band to pitch angle width and a measure of anisotropy provides additional evidence for electron scattering by whistler-mode waves. We estimate the range of resonances based on the wave properties and energies over which broadening is observed. These observations provide strong evidence that the narrowband whistler-mode waves scatter strahl-energy electrons to produce the halo and to reduce the electron heat flux.

1. Introduction

The processes that broaden strahl electrons propagating away from the Sun have been researched since early theoretical and observational studies of the solar wind (Cuperman and Harten, 1971; Hollweg, 1974; Feldman et al., 1975). Many investigations have focused on the relative roles of Coulomb collisions and interactions with plasma waves (Phillips and Gosling, 1990; Bale et al., 2013; Boldyrev and Horaites,2019). The evolution of solar wind electrons has been examined using data from satellites at radial distances from the Sun ranging from Parker Solar Probe (PSP) inside ~.2 AU (Halekas et al., 2020a, 2020b) and Helios at .3 AU (Maksimovic et al., 2005), out to Ulysses at ~4 AU (Štverák et al., 2009) and Cassini at >5 AU (Graham et al., 2017). These studies concluded that, in addition to the effects of Coulomb collisions, significant wave scattering was required. Because the strahl is a primary carrier of electron heat flux, studies of which wave mode provides the scattering are often characterized as determining the mode that controls the electron heat flux. The physics is tightly coupled because most of the proposed wave modes are excited by the heat flux and/or the temperature anisotropy.

Whistler-mode waves are frequently proposed to scatter strahl. Theoretical arguments have shown that whistlers with wave vectors parallel to the solar wind magnetic field must propagate sunward to match the resonance condition for anti-sunward strahl (Vocks et al., 2005; Saito and Gary, 2007). This constraint does not hold for obliquely propagating waves. Theoretical studies of the evolution of electrons and electron heat flux that have assumed parallel propagating whistlers have concluded that these waves cannot limit the heat flux (Kuzichev et al., 2019; Vasko et al., 2020). In contrast, Pistinner and Eichler (1998) examined the role of obliquely propagating whistlers, and concluded that these waves can control the electron heat flux. Utilizing particle-in-cell (PIC) simulations with an imposed electron heat flux, Roberg-Clark et al. (2018; 2019) also showed that large amplitude oblique whistlers could control the heat flux in high beta plasmas.

Near 1 AU, both parallel and highly oblique narrowband whistler-mode waves are observed. The parallel waves are usually low amplitude (dB/B<.02), occur primarily in the slow quiet wind, but also can occur in the faster wind (Lacombe et al. 2014; Stansby et al., 2016; Tong et al., 2019). The oblique waves are large amplitude(dB/B~.5) and occur in both fast and slow wind (Breneman et al., 2010; Cattell et al., 2020). Both obliquely propagating and parallel propagating waves are observed inside ~0.3 AU by PSP (Cattell et al., 2021; Agapitov et al., 2020). Direct comparisons of whistlers and electron distributions near 1 AU include Kajdič et al. (2016), who found that strahl electron distributions were broader when parallel propagating whistlers were observed in comparison to intervals without waves, and Pagel et al. (2007), who reported broadening of strahl when very low frequency (~3 Hz,<<$f_{ce}$) magnetic fluctuations, interpreted as broadband whistlers, occurred.

Observed whistler wave dependences on properties including parallel electron beta ($\beta_{e\parallel}$), electron temperature anisotropy and heat flux have been compared to possible instability mechanisms. The highly oblique waves observed at 1 AU by STEREO (Cattell et al., 2020; Breneman et al., 2010) occur when $\beta_{e\parallel}$ is high (between ~1 and 10), and are most consistent with the heat flux fan instability (Vasko et al., 2019), with amplitudes comparable to those found in flare simulations by Roberg-Clark et al. (2019). The most likely instability for the parallel waves is the heat flux instability (Gary et al., 1975; Abraham-Shrauner and Feldman,1977; Kuzichev et al., 2019) with enhanced growth rates associated with the electron temperature anisotropy (Stansby et al., 2016; Tong et al., 2019) and high $\beta_{e\parallel}$ (>3, Lacombe et al., 2014). Inside ~.3 AU, narrowband whistlers also occur when $\beta_{e\parallel}$ is high (~1 to 5), with the temperature anisotropy and normalized heat flux constrained, respectively, by the Gary et al. (1999) temperature anisotropy threshold and the Vasko et al.(2019) heat flux fan instability threshold (Cattell et al., 2021). Given the range of wave angles and dependences on plasma parameters, it is likely that multiple mechanisms are involved, such as the temperature anisotropy (Gary and Wang, 1996), different heat flux instabilities (Gary, 1975; López *et al,* 2020), or a combined one (Shaaban and Lazar, 2020). Vasko et al. (2020) examined the growth rates of parallel whistlers including the effects of both temperature anisotropy and heat flux. We note that, in 2-d particle-in-cell simulations initialized with core and strahl electrons based on PSP observations, Micera et al. (2020) found that initially highly oblique waves were destabilized and scattered the strahl, and subsequently parallel whistlers were excited which further scattered the strahl to produce the halo. This suggests that the variable wave angles may be related to the time evolution of the strahl along a given flux tube.

In this letter, we focus on evidence from Parker Solar Probe for scattering of strahl-energy electrons by narrowband whistler-mode waves. Using waveform capture data and electron distributions, we present the first direct evidence of this scattering at distances less than 0.3 AU from the Sun. Data sets are presented in section 2. Overviews of two intervals showing wave spectra, electric and magnetic field waveforms, and electron pitch angle data, a statistical comparison of wave power to broadening of electron pitch angle distributions, and a set of electron distributions are presented in section 3. Comparisons to previous studies, discussion and conclusions are given in section 4.

## 2. Data sets and overview

We utilize data from the Parker Solar Probe FIELDS (Bale et al., 2016) and Solar Wind Electrons Alphas and Protons Investigation (SWEAP) (Kasper et al., 2016) instrument suites. From FIELDS, we utilize the Level 2 waveform capture data for the three electric field components and the three search coil magnetic field components obtained during the first solar encounter. The waveform data were obtained for ~3.5 s intervals at ~150 ksamples/s. Storage and transmission of these data was controlled by a quality flag, and in the first encounter dust impacts often triggered the quality flag. Therefore, the waves observed in this data set may not be the largest that occur. To further examine the occurrence, duration and amplitudes of the whistler-mode wave activity, we show one electric field and one magnetic field channel in the DC coupled bandpass filter (BPF) data which is obtained at a cadence of 1 spectrum every ~1.7 s, over a frequency range from 0.4 Hz to 4000 Hz. We also utilized one electric field and one magnetic field channel in the higher frequency resolution DC coupled spectral data, which is obtained at a rate of 1 spectra every ~28 seconds, over a frequency range of ~10 Hz to 4.8 kHz (Malaspina et al., 2016). The Level 2 quasi-static magnetic field data in RTN (radial-tangential-normal) coordinates at ~4 samples per second are used for determining pitch angles and the background magnetic field structure.

The electron parameters were obtained from the SWEAP Solar Probe Analyzers (SPAN-A-E and SPAN-B-E)(Whittlesey et al. 2020). We show pitch angle distributions for energies from ~20 to 2000 eV, covering core, halo and strahl (Halekas et al., 2020a). A complete energy-angle scan is obtained in .256 Cy (~.87 s), which are summed on board to obtain a distribution every ~28 s. The solar wind velocity, used to transform distributions into the plasma frame, was obtained from the Level 2 Solar Probe Cup (SPC) moments (Case et al., 2020). For some intervals, we also show heat flux. The solar wind density, and electron core and nonthermal temperatures were obtained from the Fields Quasi-thermal Noise (QTN) data (Moncuquet et al., 2020).

## 3. Observations of whistlers and electron scattering

Figure 1 presents an overview of one interval that contains large amplitude narrrowband whistler-mode waves (2018 November 2 from 12:00 to 13:20 UT). Two different waves types are clearly distinguishable in the BBF electric field spectrum (panel a) and the BBF magnetic field spectrum (panel b). The waves at harmonics of $f_{ce}$, seen intermittently from ~11:10 to 11:50 UT and again at ~12:16 and 12:24 UT in panel a, identified as electrostatic whistler/Bernstein waves, are seen primarily in regions of quiet radial magnetic field (Malaspina et al., 2020). The electron pitch angles are narrowly field-aligned as expected for strahl, indicating that these waves do not strongly scatter electrons at these energies (panels d, e and f). Note that both the electric field spectrum and the waveform captures indicate that much weaker electron Bernstein waves occurred simultaneously with the whistlers around 1229 UT. The whistler waves, observable as peaks in the power at ~one hundred to few hundred Hz (~0.1 $f_{ce}$ ) in panels a and b from ~12:25 to ~12:37, occur when the magnetic field is more variable and usually smaller (panel c). Strong scattering at energies from ~200 to ~500 eV (panels d-f) is seen in association with the most intense whistler waves from ~1227 to ~1231 UT, and, as shown below  in Figure 4, scattering extends to ~800 eV. Scattering at the lower energies continues in concert with the weaker waves. This association provides strong evidence for scattering of strahl-energy electrons by narrowband whistlers.  Panel g, which plots a comparison of electron heat flux (black) to the Vasko et al. (2019) heat flux fan instability limit (red), is consistent with whistler regulation of the heat flux. The most intense waves are associated with increases in the core electron temperature, and increases in the plasma density.

Figure 2 presents an example of correlated electron scattering and whistler waves that occurs within magnetic field 'switchbacks' (Bale et al., 2019; Kasper et al., 2019; Dudok de Wit et al., 2020; Horbury et al., 2020) on 2018 November 3 9:00 to 11:20 UT. The signature of the switchbacks is clearly seen in panel c, which plots the radial component of the magnetic field (red) with the radial component (blue) of the ion flow (with 300 km/s to more clearly show the variations). The radial magnetic field rapidly changes from negative ~50 nT to positive values and then back, for example at ~925 UT, ~933 UT, ~939 UT and ~948 UT, correlated with an increase in the flow velocity. Note that in the region of radial field between switchbacks weaker electrostatic waves at harmonics of $f_{ce}$ (like those seen in the first ~30 minutes of Figure 1) sometimes occur.  Whistler waves (seen in the peaks in power at ~few 100 Hz in panels a and b) often fill the switchbacks, and electrons are strongly scattered. Wave amplitudes are not as large within the switchbacks as they are later in the event (~10:18 to 10:45 UT), when the five waveform captures were obtained. After ~10:45 UT, when the magnetic field is less variable, there are no whistler waves and the electrons are strongly field-aligned. In the intervals with strong whistlers, the pitch angle distributions are broadened. The normalized heat flux, on average, is smaller than in Figure 1, most often within the switchbacks when the whistlers are intense. As in the previous example, increases in the core electron temperature are observed with the most intense waves. In several cases, the temperature increased from ~25 eV to ~40 to 60 eV, suggesting that the waves may also heat core electrons.

Understanding the nature of the scattering process requires comparison of the electron distribution functions to the properties of the waveforms. The bottom panels of Figures 1(#1-#3)and 2(#2-#6) plot 0.2 s snapshots of large amplitude packets from within the 3.5 s waveform captures obtained during these two time periods. The approximate times of the snapshots are indicated by the blue triangles above panel a. Figure 2 #1, which plots the entire 3.5 s waveform capture, illustrates the wave packet structure and variability often seen in the narrowband whistlers (the associated 0.2 s snapshot is shown in panel #2). Panels #1 through #3 in Figure 1 plot one component of the electric field waveform and one of the magnetic field waveform. Panels #1 through #6 of Figure 2 plot one component of the five electric field waveforms obtained in this interval on 2018 November 3. In both Figures 1 and 2, the waveforms are coherent and large amplitude (~5-15 mV/m, and 4 − 7 nT, dB/B~0.1). The example in Figure 2 #3 is less monochromatic, and, in addition, has the signature of higher frequency waves (also seen in Figure 1 #2). The wave vectors, determined using minimum variance analysis for the largest amplitude packet in each waveform capture, ranged from ~5° to 30°. Examination of the wave angles for the entire duration of each waveform capture using singular value decomposition analysis (Santolík et al., 2003*)* indicated that there were packets with wave angles up to ~70°, close to the resonance cone, resulting in significant longitudinal electric fields, as well as a component parallel to the background magnetic field. Comparison of power in the electric and magnetic components in the BPF data, indicative of the phase velocity, also provides evidence that the wave angles are variable.

The waveform capture data in Figures 1 and 2 show that the whistler wave packets vary on sub-second time scales. The ~28s averaged spectral data often under-estimates the wave amplitudes and, as expected, miss the highly variable nature of the waves that is clear in waveforms and in the BPF data. The electron distributions have the same ~28 s cadence as the spectral data, and are, therefore, averages over regions that could include strong waves interspersed with weak or no waves. For this reason, we would not expect a one-to-one correspondence between the pitch angle broadening and wave amplitudes. In addition, electrons may have interacted with waves upstream of the observations.

Utilizing the complementary information provided by high time resolution BPF data and the better frequency resolution spectral data, and a technique developed to study whistler-mode waves in the radiation belts (Tyler et al., 2019), both the frequency and peak power of the whistler-mode waves can be more accurately determined at the higher time resolution of the BBF data. We compare these power values to two methods to assess the electron broadening. Because there are cases of very broad distributions(see examples in Figure 4, panels g and h)when the peak flux is not at 180° (away from the Sun), the pitch angle width was defined to be the width at half maximum between the maximum flux (using the actual pitch angle measurement for non-180° peaks or the splined value at 180° pitch angle if the peak flux is at 180°) and the minimum flux. Note that the plotted values are full-widths. The second method is an 'anisotropy,' defined to be the measured maximum flux over the minimum flux between 90° and 180° degrees. When the strahl is narrowly peaked, the 'anisotropy' is large; when the distribution is very broad, the anisotropy is very small. Figure 3 plots the pitch angle width and 'anisotropy' for the energy band centered at 314 eV versus the magnetic power ($nT^2$/Hz) in the whistler-mode frequency band for the entire two day period containing the shorter intervals shown in Figures 1 and 2. The median value for the power is indicated by the red squares; the upper quartile (75%) and lower quartile ( 25%) are indicated by the red diamonds. It is clear that the narrowest pitch angle distributions (<40° full width, or 20° half width), corresponding to large 'anisotropies,' are on average associated with the lowest wave power. The broadest distributions (large pitch angle width and small anisotropy) are clearly correlated with the largest amplitude waves. This provide strong statistical evidence for electron scattering by the narrowband whistler-mode waves.

Energy-pitch angle distributions of the electrons provide more detailed diagnostics for the interactions with the whistler waves. Figure 4 presents examples of the types of distributions observed. For these events, a pitch angle of 0° refers to electrons traveling along the magnetic field towards the sun, and 180° refers to electrons traveling along the magnetic field away from the sun. Times given are the center time of the 28 s interval. Energies below 20 eV are not included due to possible secondary electron effects (Whittlesey et al.2020; Halekas et al., 2020a). Panels a and b show narrow, field-aligned distributions streaming away from the Sun over energies of ~100 eV to 1 keV, consistent with strahl. Comparison to Figure 1 (panels a and b) shows that no whistler-mode waves were observed at the time of the distribution in Figure 4 panel b, and very weak waves at the time of the distribution in Figure 4 panel a (times indicated by black triangles below Figure 1a). The broadening of the pitch angles seen in panels c, d and e (obtained at the times indicated by the two yellow triangles in Figure 2 and the yellow triangle in Figure 1)is energy-dependent, primarily between ~250 and 600 eV. The case in panel e is broader at low energies. In contrast, the cases in panels f, g and h (obtained respectively at the time indicated by the second red triangle in Figure 1, the red triangle in Figure 2, and the first red triangle in Figure 1) have distributions that are broadened over the range from < 100 eV to ~1 keV, although the extent of the broadening varies significantly. The extreme broadening seen in Figure 4h (2018 November 2 12:30:11 UT) which extends to lower energies is not a unique case; at least two other distributions (12:25:03 UT, 12:27:51 UT) are nearly identical. The three waveform captures (Figure 1) were obtained at this time, and the large amplitudes continue for adjacent times as can be inferred from the BPF data. Similarly the very broad distribution in Figure 4g on 2018 November 3 is similar to ones at 10:15:04 UT and 10:43:58 UT. There are a number of cases with peaks that are not at 180° in some energy bands (seen most clearly in panels g and h, but also observable at high energies in d and e). Rotations of the magnetic field during the sampling interval could lead to a peak not at 180°, but we have carefully selected distributions obtained when field direction was stable. Similar peaks at an angle to the magnetic field are seen in the results from our particle tracing code (Vo et al., 2020).

Table 1 presents a qualitative relationship between the whistler-mode wave amplitudes and durations and the twenty-six representative energy-pitch angle distributions (categorized as narrow, broadened and very broad, as discussed above) that were obtained when the distributions were not affected by the rapid changes in the magnetic field (13 on 2018 November 2, and 13 on 2018 November 3). As discussed above, a perfect correlation would not be expected because the obliquity of the waves varies, the electrons may have interacted with waves upstream of the observation location, and due to the mismatch between wave packet and distribution function timescales. Waves at the time of each distribution were categorized as: (A) very intense (BPF electric field amplitude >4 x$10^{-4}$ V, BPF magnetic field >.05 nT, spectral electric field >1x$10^{-8}$ V$^2$/Hz and spectral magnetic field >5.0x$10^{-4}$ nT$^2$/Hz, in the appropriate frequency band); (B) intense (meeting criteria for very intense for only part of the interval, or only for the BPF or the spectra, or occurring in all channels at reduced intensity); (C) moderate (occurring for part of interval with reduced intensities(BPF electric field amplitude >1.5 x10$^{-4}$ V, BPF magnetic field >.015 nT, spectral electric field >1x10$^{-9}$ V$^2$/Hz and spectral magnetic field >1.0x10$^{-5}$ nT$^2$/Hz, in the appropriate frequency band); (D) waves detected only in the electric field; and (E) very weak intermittent waves or no detectable waves. Table 1 summarizes the comparison of distribution characteristics and wave amplitudes. Of the two narrow distributions, the narrowest, Figure 4b, occurred when the waves were below the threshold (E), and Figure 4a occurred when the waves were very intermittent (E). Distributions that have evidence of pitch angle scattering usually occur with moderate or large whistler waves; fifteen broad or very broad distributions are simultaneous with intense (B) or very intense (A) whistlers, and five with moderate (C) waves. These comparisons of individual distributions to wave amplitudes complement and reinforce the conclusion from Figure 3 that the whistler-mode waves scatter electrons, and indicate that the scattering can occur over a broad energy range. Comparison of distributions that are closely spaced in time suggests that increases in pitch angle may sometimes be associated with decreases in energy.

4. Discussion and conclusions

The intervals shown above provide strong evidence for scattering of solar wind electrons at energies of ~100 eV to ~1 keV by narrowband whistler-mode waves to produce the halo and reduce the electron heat flux. This is consistent with the results of a particle tracing code (Vo et al., 2020; Breneman et al., 2010) for wave and plasma parameters based on 1 AU measurements. Strong scattering by oblique whistlers has been observed in PIC simulations of the solar wind (Micera et al., 2020), and in other contexts including solar flares (Roberg-Clark et al., 2019), the radiation belts (Katoh and Omura, 2007), and high beta astrophysical plasmas (Roberg-Clark et al., 2018). PIC simulations have also provided evidence for scattering by parallel whistlers in the radiation belts (Camporeale and Zimbardo, 2015), for the case of sunward propagating waves in the solar wind (Saito and Gary, 2007), and during later times in the simulation of Micera et al. (2020).

The intense narrowband whistlers occur primarily in regions where the solar wind magnetic field is variable and often weaker, and can occur in association with magnetic switchbacks (Agapitov et al., 2020). Statistics on narrowband whistlers inside ~.3 AU (Cattell et al., 2021) confirmed this association. Figure 2 illustrated that although the whistlers are strongest near the switchback edges, intermittent weaker waves fill the entire switchback. The occurrence of intense whistlers, electron scattering and low normalized heat flux within switchbacks may aid in understanding their origin.

The events we examined occur when solar wind speeds were ~250 to ~350 km/s. Several studies have examined differences in the evolution of the non-thermal electron distributions between the slow and fast wind. Pagel et al. (2005) suggest that different scattering mechanisms are active in the fast and slow wind, based on differences in the energy dependence of pitch angle widths observed at 1 AU. Stverak et al. (2009) noted differences in the evolution with radial distance for non-thermal electrons for fast and slow wind, and suggested that Coulomb collisions were less important in the fast wind. The narrowband whistlers occur primarily in regions of high beta both inside .3 AU and at 1 AU (Cattell et al., 2020; 2021). In a study of strahl evolution between .3 AU and 1 AU, Berčič et al. (2019) showed differences in scattering for high versus low core electron beta. For high beta, strahl electrons were scattered over the full energy range, which they suggest is due to stronger kinetic instabilities. This is consistent with our observations that strong scattering often occurs over a broad energy range.

For the set of 15 waveform captures obtained on 11/2/2018 and 11/3/2018, we calculated the range of resonant energies for the observed wave frequencies, wave angles, and wave vector magnitude, using the resonance condition, $\omega - \vec{k} \cdot \vec{v_e} = n\Omega_e$, where $\omega$ is the wave frequency, $\vec{k}$ is the wave vector, $\vec{v_e}$ is the electron velocity and $\Omega_e$ is the electron gyrofrequency. We include the Doppler shift of the wave frequency and the electron velocity relative to the solar wind flow. Using the properties determined for the largest amplitude packet within each event, the ratio of our measured frequencies to the electron gyrofrequency is ~0.09 to ~0.3. The phase velocities ranged from ~650 km/s to ~1100 km/s, larger than the solar wind speeds of ~250 to 350 km/s. The wave angles varied from ~3° to ~45°, and the magnitude of the wave vector varied from ~.6 to 2.5 km$^{-1}$. Resonant energies ranged from ~100 eV to 700 eV for the n=+1 resonance, and ~90 eV to ~400 eV for the n=-1 resonance (for sunward propagating waves). The range of energies over which the electron distributions are broadened is consistent with the inferred resonant energies. For the Landau resonance, $\frac{\omega}{k_\parallel} = v_{e\parallel}$, the resonant energies range from a few to 10s of eV, which might be associated with the increase in core electron temperature seen in association with the largest whistlers. This core heating was also observed in a statistical study of whistler waves observed by PSP (Cattell et al., 2021), and in the particle tracing simulations (Vo et al., 2020).

For the three cases (11/2/2018 12:27:51 and 12:30:11 and 11/3/2018 10:43:58) where a 3.5 s waveform capture was obtained within the 28 s interval of one of the 26 electron distributions, we can use the observed energies over which broadening occurred to estimate the range of resonance number. There are two issues to consider in determining the range of resonances involved: the waves occur in packets with a range of frequencies; and, over the 28 seconds needed to obtain the distribution function, multiple wave packets with varying wave angles may occur. For example, in the 3.5 s of the capture at 10:43:58 on 11/3/2018, wave angles for the packets varied from ~8° to 30°, and in the 10:30:31 capture, angles varied from ~30° to ~70°. The energy associated with n=1 was ~100 eV, and the n=3 resonance reached energies of 900 eV to ~1 keV, consistent with the energies over which broadening occurred (from ~100 eV to ~1 keV). Note that Roberg-Clark et al. (2019) found scattering consistent with resonances higher than n=1 in PIC simulations of whistlers.

With the assumption that the waves are parallel propagating, many authors have shown that scattering rates are proportional to $\Omega_e \delta B^2 / B^2 \propto \delta B^2 / B$ (Brice, 1964; Kennel and Petschek, 1966; and Albert, 2017, for whistlers in the context of radiation belt). An analogous argument can be made for oblique waves by transforming into the de Hoffmann-Teller frame rather than the wave frame, because in this frame the wave propagates along the magnetic field. This ratio is largest for the waveform captures at 10:19:15 and 10:43:47 on 11-3-2018 (both close to very broad distributions), and somewhat smaller for 10:30:29 on 11-3-2018 and 12:27:57 *and* 12:30:05 on 11-2-2018 (all near broad distributions). Although this is not a statistically significant number of events, it is consistent by the statistical results for $\delta B^2 / B$ obtained as for the power in Figure 3 (not shown).

We show directly, for the first time, that the strahl energy electrons, which carry the heat flux, are strongly scattered by the narrowband whistler-mode waves, reducing the heat flux in the solar wind inside .3 AU. These narrowband large amplitude whistler-mode waves are, therefore, the most likely candidates for regulating the electron heat flux and scattering of strahl electrons into the halo, as confirmed by PIC simulations (Micera et al., 2020) and the particle tracing results (Vo et al., 2020).

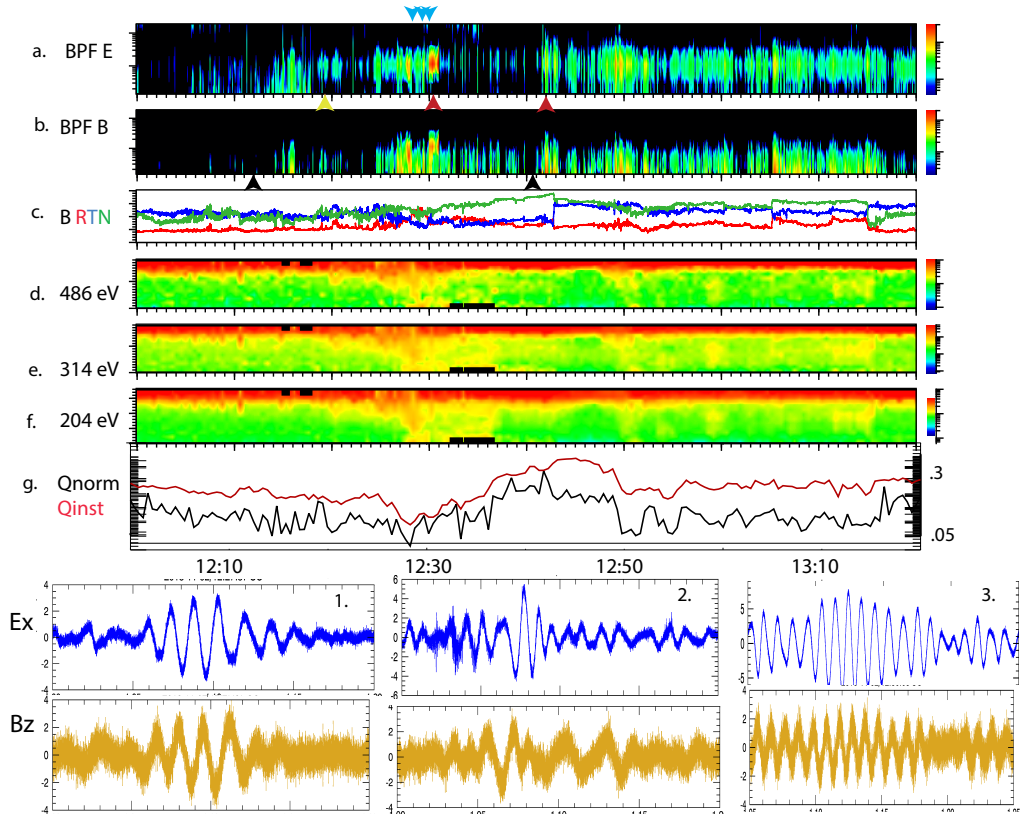

Figure 1. November 2, 2018 from 12:00 to 13:20. From top to bottom: DC coupled electric field BPF spectrum from 4 Hz to 4 kHz; DC coupled magnetic field BBF spectrum from 4 Hz to 4 kHz; the magnetic field in RTN coordinates (R in red, T in blue and N in green); pitch angle spectra for electrons with center energies of 486 eV, 314 eV, and 205 eV; normalized heat flux(black) and heat flux fan instability threshold from Vasko et al. (2019) (red). Bottom panels: 0.2 s snapshots from the three waveform captures obtained during this interval at the times of the blue triangles, Ex (blue) and Bz (gold).

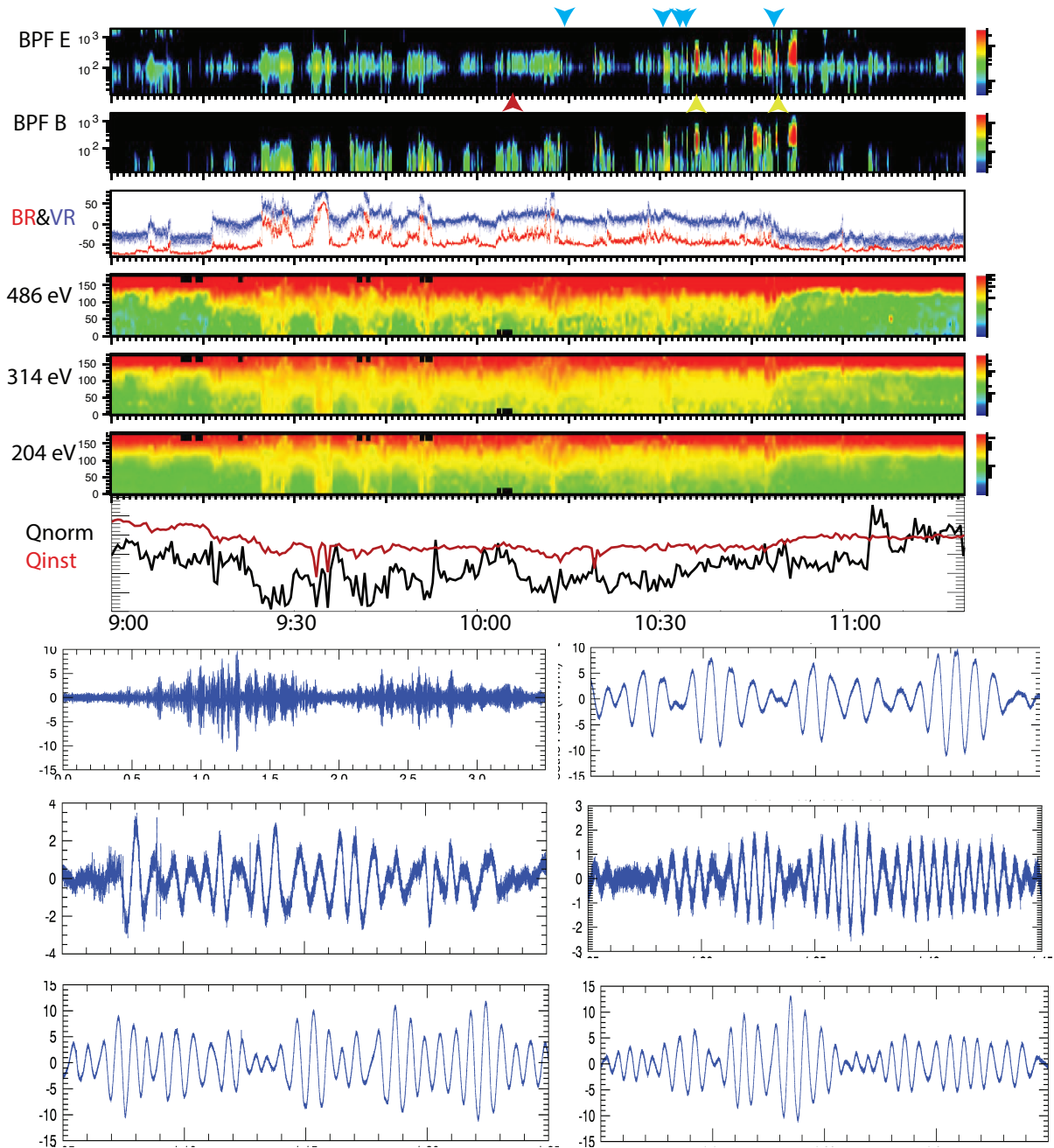

Figure 2. Parker Solar Probe observations of waves, solar wind and electron distributions. From top: DC coupled electric field and search coil magnetic field BPF spectra (12 Hz to 4000 Hz); R(radial) component of the magnetic field (red) with the R component of the ion flow velocity (with 300 km/s subtracted) in blue; pitch angle spectra for electrons with center energies of 486 eV, 314 eV, and 204 eV; normalized heat flux (black) and fan instability threshold from Vasko et al. (2019) (red). Bottom panels plot waveforms at the times indicated by the blue arrows: #1 plots one full 3.5 s waveform, and #2–#6 plot 0.2 second snapshots of the SC-X component of the electric field from the five 3.5 s burst waveform captures transmitted during this interval.

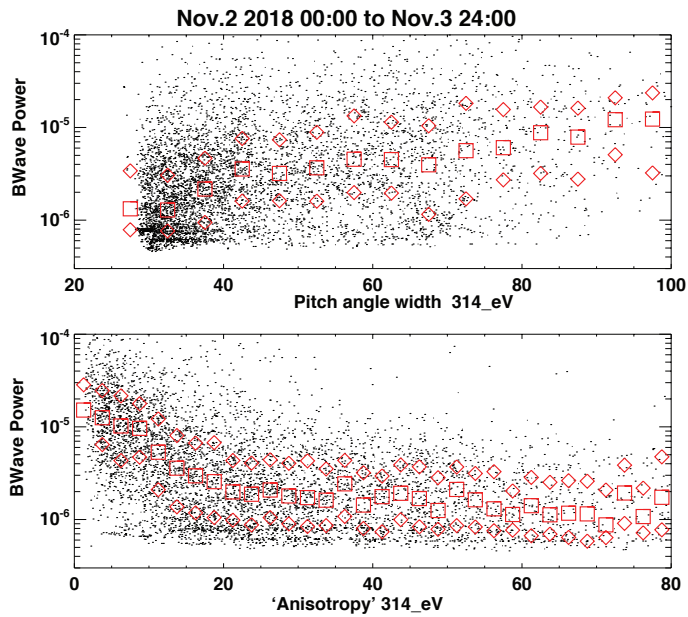

Figure 3. Scatter plot of whistler magnetic field wave power in $nT^2/Hz$ in the whistler band versus pitch angle width and 'anisotropy'- defined as the measured maximum flux over the measured minimum flux between 90 degrees and 180 degrees for the two days containing the time intervals in Figure 1 and 2. The median values of the wave power are over-plotted as red squares, and the upper 75% and lower 25% are are over-plotted as red diamonds.

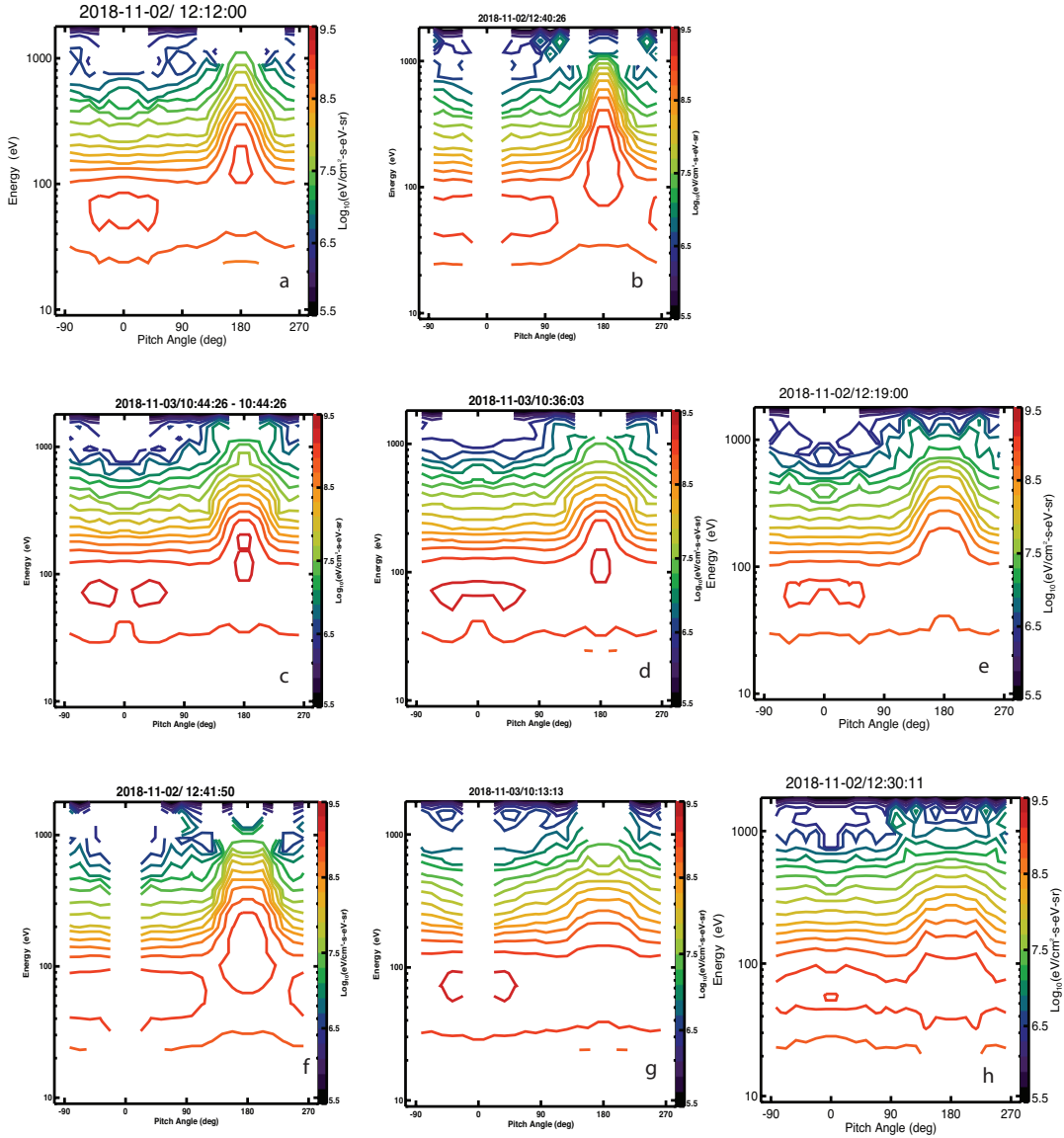

Figure 4. Examples of types of electron energy-pitch angle distributions observed during the intervals shown in Figures 1 and 2. Narrow strahl distributions are plotted in panels a and b(obtained at times of black triangles in Figure 1). Energy dependent broadening panels examples are shown in c, d and e (obtained at times of the two yellow triangles in Fig.2 and the one in Fig.1). Panel f (obtained at time of second red triangle in Fig.1)shows energy independent broadening. Panels g and h (obtained at times of red triangles in Fig.2 and first red triangle in Fig. 1)show extreme broadening with peaks off 180 degrees.

| Wave DF | E | D | C | B | A | (A+B) |
|---|---|---|---|---|---|---|
| narrow | 2 | | | | | |
| broad | | 4 | 2 | 6 | 5 | 11 |
| Very broad | | | 2 | 2 | 4 | 6 |

Table 1. Qualitative comparison of electron distribution and wave properties. Narrow, broad and very broad refer to electron pitch angle widths, as described in the text. A,B,C,D and E refer to wave characteristics, with A being the most intense, D having only electric field signatures, and E having very weak and intermittent or no waves. See text for details.

Acknowledgements: We acknowledge the NASA Parker Solar Probe Mission, the FIELDS team led S. D. Bale, and the SWEAP team led by J. Kasper for use of data. The FIELDS and SWEAP experiments on the Parker Solar Probe spacecraft were designed and developed under NASA contract NNN06AA01C. Data analysis at was supported under the same contract.